\documentclass{aa}
\usepackage{psfig}
\usepackage{graphics}
\def\lesssim{\mathrel{\hbox{\rlap{\hbox{\lower4pt\hbox{$\sim$}}}\hbox{$<$}}}}
\begin{document} 
\thesaurus{03(12.12.1; 11.03.1; 11.19.7)}
\title{Finding Galaxy Clusters using Voronoi Tessellations} 
\author{ Massimo Ramella\inst{1}
\and Walter Boschin\inst{2}
\and Dario Fadda\inst{3}
\and Mario Nonino\inst{1}}
\offprints{M. Ramella (ramella@ts.astro.it)}
\institute{Osservatorio Astronomico di Trieste, Via Tiepolo 11, I-34100 - Trieste, Italy; ramella@ts.astro.it, nonino@ts.astro.it;
\and Dipartimento di Astronomia, Universit\`{a} degli Studi di Trieste, Via Tiepolo 11, I-34100 - Trieste, Italy; boschin@ts.astro.it;
\and Instituto de Astrofisica de Canarias, Via Lactea S/N, E-38200 - La Laguna (Tenerife), Spain; fadda@ll.iac.es}
\date{Received xx xxxx 2000 / Accepted xx xxxx 200x}
\authorrunning{M. Ramella et al.} 
\titlerunning{Finding Galaxy Clusters using Voronoi Tessellations}
\maketitle

\begin{abstract}

We present an objective and automated procedure for detecting clusters
of galaxies in imaging galaxy surveys\footnote{The code described in
this paper is available on request}. Our Voronoi Galaxy Cluster Finder
(VGCF) uses galaxy positions and magnitudes to find clusters and
determine their main features: size, richness and contrast above the
background.  The VGCF uses the Voronoi tessellation to evaluate the
local density and to identify clusters as significative density
fluctuations above the background. The significance threshold needs to
be set by the user, but experimenting with different choices is very
easy since it does not require a whole new run of the algorithm. The
VGCF is non-parametric and does not smooth the data.  As a
consequence, clusters are identified irrispective of their shape and
their identification is only slightly affected by border effects and
by holes in the galaxy distribution on the sky.  The algorithm is
fast, and automatically assigns members to structures.

A test run of the VGCF on the PDCS field centered at $\alpha = 13^{h}$
26$^{m}$ and $\delta$ = +$29^{o}$ 52' (J2000) produces 37 clusters. Of
these clusters, 12 are VGCF counterparts of the 13 PDCS clusters
detected at the 3$\sigma$ level and with estimated redshifts from
$z=0.2$ to $z=0.6$.  Of the remaining 25 systems, 2 are PDCS clusters
with confidence level $< 3\sigma$ and redshift $z \leq 0.6$.
 
Inspection of the 23 new VGCF clusters indicates that several of these
clusters may have been missed by the matched filter algorithm for one
or more of the following reasons: a) they are very poor, b) they are
extremely elongated, c) they lie too close to a rich and/or low
redshift cluster. 

\end{abstract}


% Section 1

\section{Introduction}

Wide field imaging is becoming increasingly common since new large
format CCD cameras are, or soon will be available at several telescope
(see e.g. MEGACAM, Boulade 1998, Boulade et al. 1998, WFI, Baade 1999,
etc).  The possibility to perform wide field imaging of the
extragalactic sky allows a systematic search of medium-high redshift
galaxy clusters in two-dimensional photometric catalogs of galaxies.
These candidate clusters are of cosmological interest and are primary
targets for subsequent follow-up spectroscopical observations (see,
e.g., Holden et al. 1999, Ramella et al. 2000).

Several automated algorithms have already been developed for the
detection of clusters within two dimensional galaxy catalogs.  The
classical techniques used for this task are the ``box count'' (Lidman
\& Peterson 1996) and the ``matched filter'' algorithm proposed by
Postman et al. (1996, hereafter P96) and its recent refinements (see
Kepner et al., 1999, Kawasaki et al. 1998, Lobo et al 1999).

The box-counting method uses sliding windows (usually squares) which
are moved across the point distribution marking the positions where
the count rate in the central part of the window exceeds the value
expected from the background determined in the outermost regions of
the window.  The main drawbacks of the method are the introduction of
a binning to determine the local background, which improves count
statistics at the expense of spatial accuracy, and the dependence on
artificial parameters like bin sizes and positions or window size and
geometry.

The ``matched filter'' is a maximum-likelihood (ML) algorithm which
analyzes the galaxy distribution with the assumption of some model
profiles to fit the data (e.g. a density distribution profile and a
luminosity function). This last technique has been used to build the
Palomar Distant Cluster Survey (PDCS, see P96) catalog and the EIS
cluster catalog (Olsen et al. 1999, Scodeggio et al. 1999). These two
catalogs are two of the largest presently available sets of distant
clusters, with 79 and 302 candidate clusters respectively.

However, the main drawback of the matched filter method is that it can
miss clusters that are not symmetric or that differ significantly from
the assumed profile.  This can be a serious problem since we know that
a large fraction of clusters have a pronounced ellipticity (see
e.g. Plionis et al. 1991, Struble \& Ftaclas 1994, Wang \& Ulmer 1997,
Basilakos et al.  2000). Furthermore, the matched filter technique is
sensitive to border effects because the ML is computed on a circular
area. This is a problem because real galaxy catalogs are finite and
usually present holes in regions corresponding to camera defects or
bright stars.

Other interesting methods to detect overdensities in a galaxy catalog
are the LRCF method (Cocco \& Scaramella 1999), kernel based
techniques (Silvermann 1986, Pisani 1996), and wavelet transforms
(Bijaoui 1993, Fadda et al. 1997). In general, all these techniques
are very sensitive to symmetric structures and suffer from border
effects.

Ebeling \& Wiedenmann (1993) develop a method based on Voronoi
tessellation (VTP) in order to identify overdensities in a Poissonian
distribution of photon events. Their goal is to detect sources in
ROSAT x-ray images. The VTP does not sort points into artificial bins
and does not assume any particular source geometry for the detection
process. These interesting properties are also very attractive for
other applications. For example, Meurs and Wilkinson (1999) present an
application of VTP to a slice of the CfA redshift survey with the
purpose of isolating the bubbly and filamentary structure of the
galaxy distribution. Several other applications of the Voronoi
tessellation to astronomical problems have been published (see, e.g.,
van de Weygaert 2000, El-Ad \& Piran 1996, 1997, Ryden 1995, Goldwirth
et al. 1995, Zaninetti 1995, Doroshkevich et al. 1997, Ikeuchi \&
Turner 1991, Icke \& van de Weygaert 1991). Clearly, Voronoi
tessellation is also interesting for the search of galaxy clusters
(Ramella et al. 1999, Kim et al. 2000).

In this paper we present an automatic procedure for the identification
of clusters within two-dimensional photometric catalogs of
galaxies. Our technique, announced in Ramella et al. 1999, is based on
the VTP of Ebeling \& Wiedmann (1993). The procedure, hereafter
Voronoi Galaxy Cluster Finder (VGCF), is completely non-parametric,
and therefore sensitive to both symmetric and elongated and/or
irregular clusters. Moreover, the VGCF is not significantly affected
by border effects, and automatically assigns members to structures.
\begin{figure}[t]
\psfig{figure=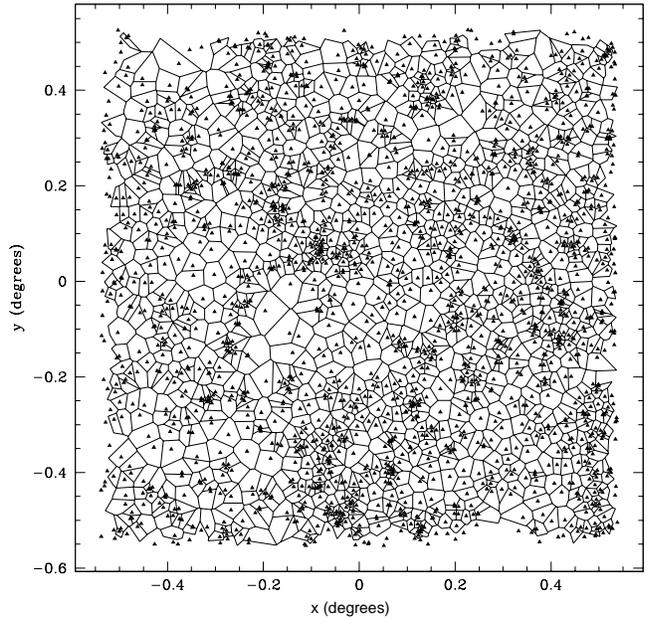,width=9cm,angle=0}
\caption{Voronoi tessellation of a galaxy field.}
%\label{fig1}
\end{figure}

\begin{figure}[b]
\psfig{figure=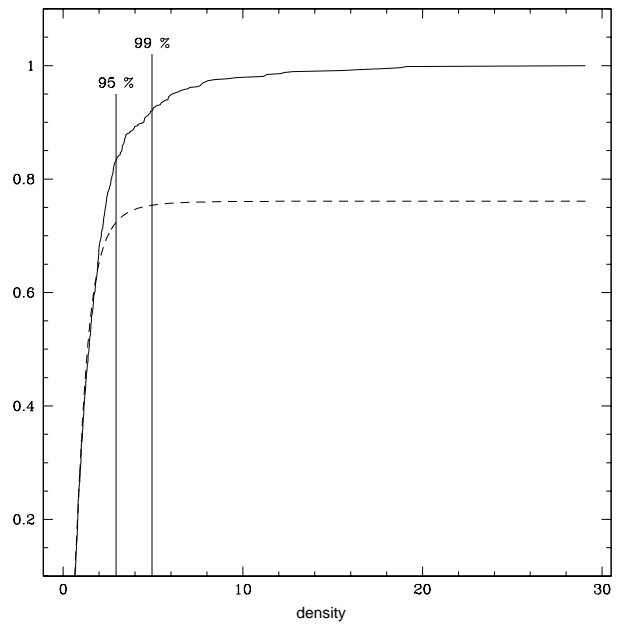,width=9cm,angle=0}
\caption{Density distribution for the points in Fig. 1 (solid line)
and the fitted Kiang distribution (dashed line). The two vertical
lines correspond to the 95\% and 99\% density thresholds for the
selection of clusters.}
%\label{fig2}
\end{figure}

In Sect. 2 we describe the method and in Sect. 3 we show an
application of the VGCF to the $\alpha = 13^{h} 26^{m}, \delta =
+30^{o}52^{'}$ field of the Palomar Distant Cluster Survey (see
P96). In particular, we illustrate the performances of the algorithm
and discuss some aspects of the its application that are specific to
the construction of a catalog of galaxy clusters.  Finally, in Sect. 4
we summarize our results.

In what follows we assume an Hubble constant $H_{0}=100\, km\, s^{-1}
\,Mpc^{-1}$ and a deceleration parameter $q_{0}=1/2$.

% Section 2

\section{The method}

A Voronoi tessellation on a two-dimensional distribution of points
(called nuclei) is a unique plane partition into convex cells, each of
them containing one, and only one, nucleus and the set of points which
are closer to that nucleus than to any other (see Fig. 1).  For us,
any point is a galaxy of a given galaxy catalog.  The algorithm used
here for the construction of the tessellation is the ``triangle'' C
code by Shewchuk (1996).  In order to avoid border effects, we
restrict our study to Voronoi cells which do not intersect the convex
hull of the set of points. In the case of galaxy catalogs with holes,
we do not consider Voronoi cells that intersect the hole borders. To
do that, we model the hole with a combination of rectangles and we
reject cells which have at least one vertex inside one or more of
these rectangles.
\begin{figure}[t]
\psfig{figure=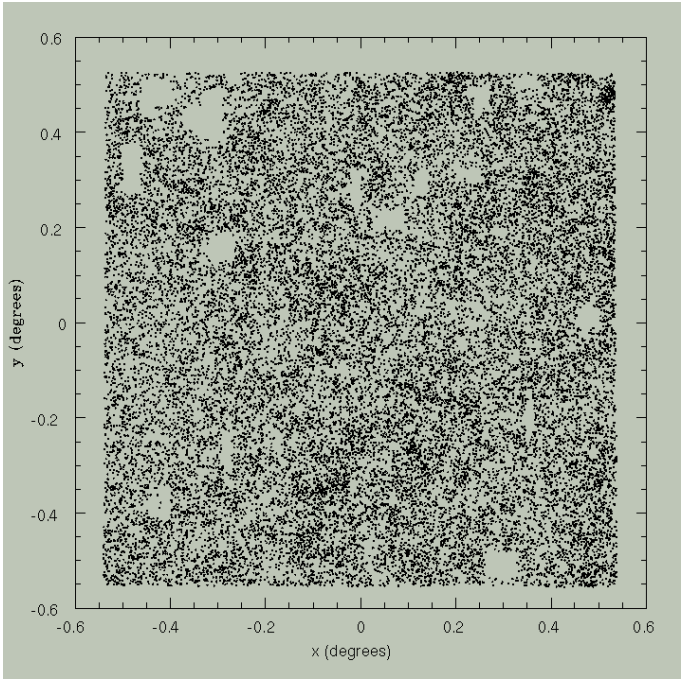,width=9cm,angle=0}
\caption{The 25432 galaxies of the PDCS field centered at $\alpha=$
201.5 degrees and $\delta=$ 29.9 degrees (J2000). Holes correspond to
bright stars in the field. North is on the top and East on the left of
the figure.}
%\label{fig3}
\end{figure}

Since we want to find overdensities in a catalog of galaxies, the
interesting quantity for us is the density, i.e. the number of
galaxies for area unit, rather than the cell area associated by the
Voronoi tessellation to each galaxy. An empirical distribution for
randomly positioned points following Poissonian statistics, has been
proposed by Kiang (1966):
\begin{equation} 
dp(\tilde{a})=\frac{4^4}{\Gamma(4)}\tilde{a}^3 e^{-4\tilde{a}}d\tilde{a},
\end{equation} 
where $\tilde{a}\equiv a/<a>$ is the cell area in units of the average
cell area $<a>$. As each cell contains exactly one galaxy the
corresponding density is the inverse of the cell area $f\equiv 1/a$.
The idea of Ebeling \& Wiedenmann (1993) is to estimate the background
by fitting the Kiang cumulative distribution to the empirical
cumulative distribution resulting from the real catalog in the region
of low density which is not affected by the presence of structures
($\tilde{f}\equiv f/<f>\,\le0.8$). We can thus establish a density
threshold (Fig. 2) and define as overdensity regions those composed by
adjacent Voronoi cells with a density higher than the chosen
threshold.

In this paper, the density threshold is to separate background regions
and fluctuations which are significant overdensities at the 80\%
c.l.. Since some of the selected overdensities can still be random
fluctuations rather than physical clusters, it is necessary to further
suppress random overdensities in order to save only significant
clusters. We proceed as follows: we compute the probability that an
overdensity corresponds to a background fluctuation by:
\begin{equation} 
N(\tilde{f}_{min},\le n) = N_{bkg}\,\frac{\,n_{src,fluc} (\tilde{f}_{min},0)
\,e^{-b\tilde{f}_{min}n}}{b(\tilde{f}_{min})}, 
\end{equation} 
where $n$ is the number of points associated to the cluster, $N_{bkg}$
is the number of background galaxies, and $\tilde{f}_{min}$ is the
minimum normalised density of the cells belonging to the cluster. Both
$n_{src,fluc}$ and $b$ are functions of $\tilde{f}_{min}$, which can
be computed from simulations (Ebeling \& Wiedemann, 1993).  We reject
overdensities whose probability to be a random fluctuation of the
background is greater than 5\%.
\begin{figure}[t]
\psfig{figure=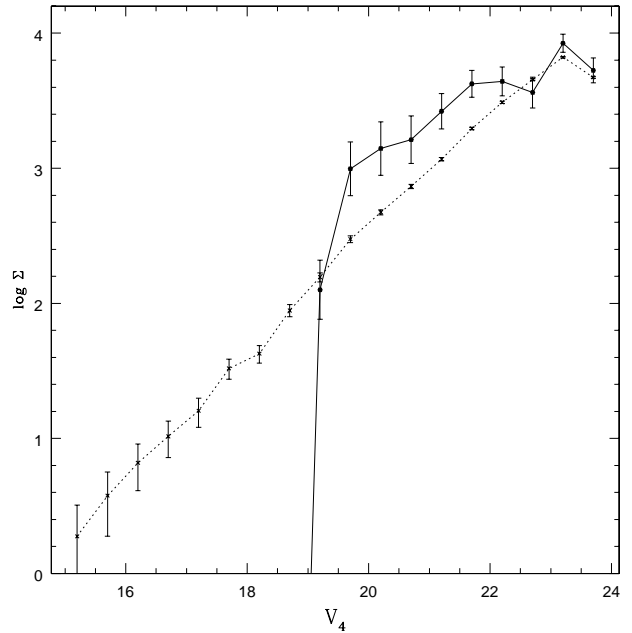,width=9cm,angle=0}
\caption{The solid line is the differential magnitude distribution of
galaxies of the PDCS cluster n. 51 ($z_{PDCS}=0.4, N_R=80$). The
dotted line is the magnitude distribution of average galaxy
counts. Counts are normalized to an area of one square degree.}
%\label{fig4}
\end{figure}

Once we detect significant clusters, we regularize their shape. We
assume that the points inside the convex hull defined by the set
of points belong to the cluster itself. Then we fit a circle to each
cluster and we expand it until the mean density inside the circle is
lower than the density of the original cluster. We perform this
expansion of the original cluster because the external points of a
cluster have usually large area cells, typical of the low density
background. These cells can not be associated from the beginning to
the cluster by the Voronoi tessellation algorithm.

We stress that fitting a circle to the Voronoi cluster has absolutely
no part in the detection of the cluster. It only provides a convenient
way to catalog the cluster with a center and a radius. In principle,
we could fit an ellipse rather than a circle, since our cluster finder
is sensitive to elongated clusters. In practice this is not convenient
because, as Ebeling (1993) shows, fitted ellipse eccentricities become
insignificant for values lower than about 0.7.
\begin{figure}[t]
\psfig{figure=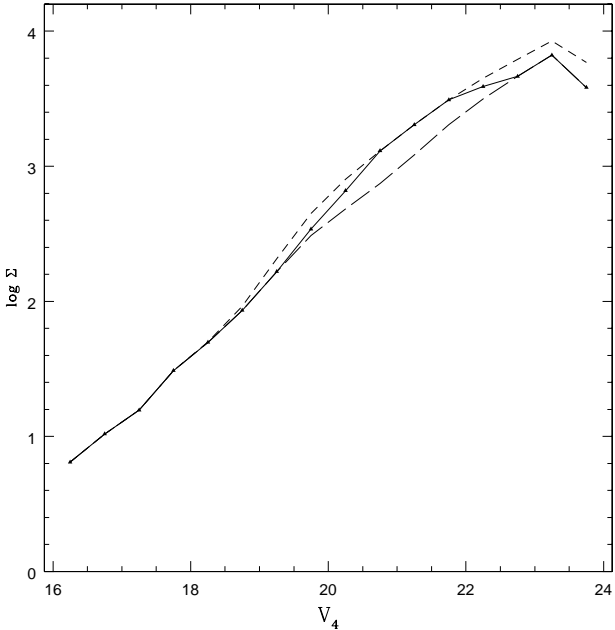,width=9cm,angle=0}
\caption{Differential magnitude distribution of galaxies in a circle
centered on a simulated cluster at redshift 0.3 and $N_{R}=60$ (solid
line). The long dashed line is the magnitude distribution of
background galaxies and the short dashed line is the magnitude
distribution of cluster galaxies without considering surface
brightness selection effects. Counts are normalized to an area of one
square degree.}
%\label{fig5}
\end{figure}
\begin{figure}[t]
\psfig{figure=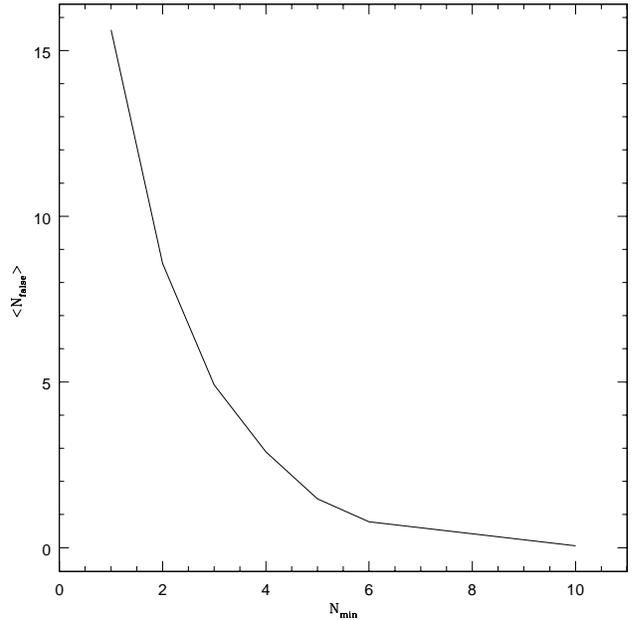,width=9cm,angle=0}
\caption{Average number of false detections per field as a function of
the threshold in number of coincident fluctuations, $N_{min}$.}
\label{fig6}
\end{figure}

At this point we have a list of clusters with position, number of
members associated (we statistically subtract the background),
contrast against the background, and projected size on the sky. We
also have a file of galaxies with position, magnitude, area of the
Voronoi tessel, and a label for appartenence to one of the detected
clusters. As a final (optional) step, we can assume a density model
and fit it on the galaxy distribution at the position where the
cluster has been found.  In this way we can have a better estimate of
the number of members and of the sizes of the cluster, in case
clusters are well described by the model.

% Section 3

\section{Building a cluster catalog: the case of a PDCS field}

In this section we test the VGCF on part of the galaxy catalog used by
P96 to produce the Palomar Distant Cluster Survey (PDCS).

In particular, we run the VGCF on the galaxy catalog derived from the
$V_{4}$ band image of the PDCS field centered at $\alpha = 13^{h}$
26$^{m}$ and $\delta$ = +$29^{o}$ 52' (J2000). The galaxy catalog
contains 25432 galaxies.  For each galaxy the catalog lists right
ascension and declination, and total magnitude in the $V_{4}$ band
(see P96, Sect. 2). The effective sky area covered by the field is
1.062 square degrees (see Fig. 3).

We also have a catalog of bright stars areas to be avoided by the 
VGCF. As already pointed out, the VGCF can very easily deal with
``holes'' in the galaxy catalog.

\subsection{Magnitude bins}

The PDCS galaxy catalog is deep, i.e. the number density of galaxies
is high. It is therefore \textit{a priori} possible that only the
highest number density contrasts survive the ``dilution'' caused by
foreground/background galaxies. For this reason we decide to run the
VGCF in magnitude bins. As we will show, to run the algorithm in
magnitude bins also helps in further suppressing spurious clusters.
\begin{figure}[!h]
\psfig{figure=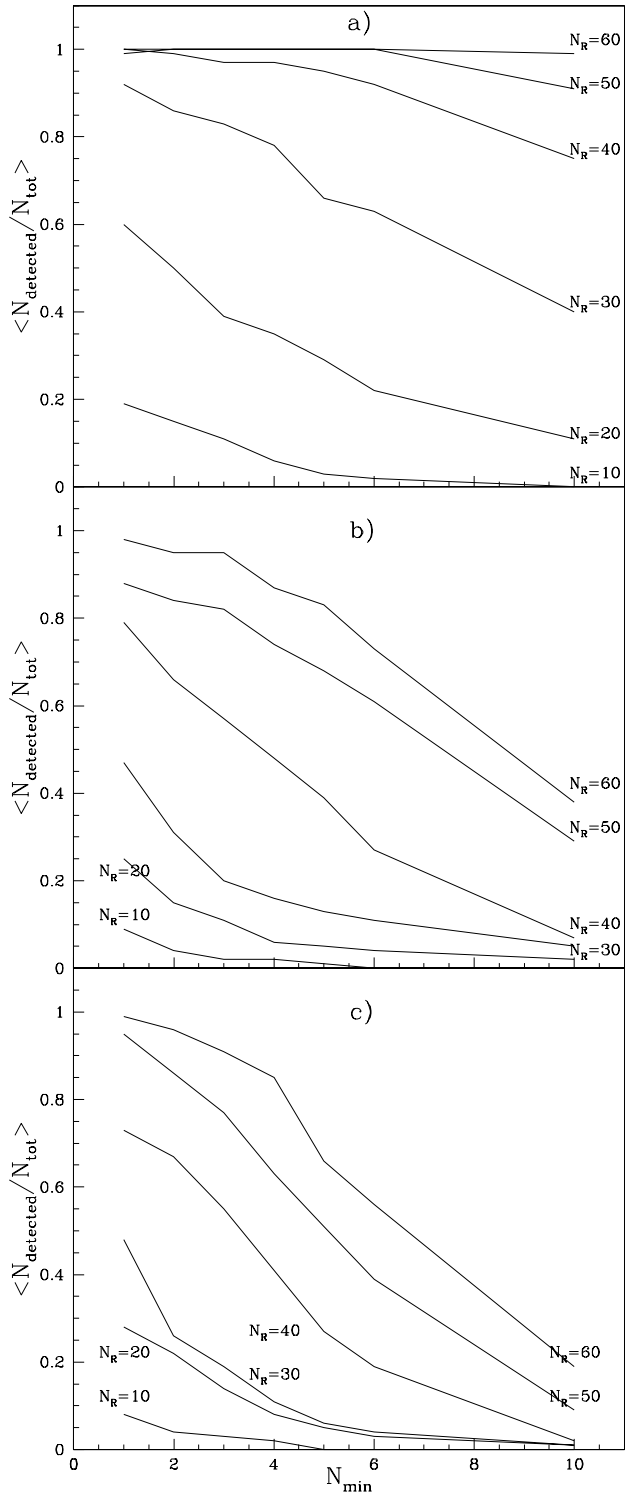,width=20cm,angle=0}
\caption{Average fraction of detections of simulated clusters at z=0.3
(panel {\it a}), 0.5 (panel {\it b}) and 0.8 (panel {\it c}) as a
function of $N_{min}$. The different curves in each panel correspond
to different values of the richness parameter, $N_R$.}
%\label{fig7}
\end{figure}

The choice of the bin size is somewhat arbitrary and cannot be equally
optimal for the detection of all clusters. In fact, the magnitude
range of the excess counts produced by clusters varies with the
cluster distance and richness. We choose to optimize the bin size for
the detection of PDCS-like clusters at intermediate redshift ($z \sim
0.3\, -\,0.6$), i.e around the peak of the galaxy selection
function.

In Fig. 4 we plot the magnitude distribution (normalized to unit
area) of galaxies within the projected area of P51 ($z_{PDCS} \sim
0.4$ and richness $N_{R} \sim 80$, see P96 for the definition of
$N_{R}$), superimposed on the average counts per square degree of the
whole PDCS field. From inspection of Fig. 4 it appears that a
reasonable choice for the bin size is two magnitudes.  With such a bin
most of the counts of the cluster peak fall within one bin and the
cluster detection is optimal. In fact, a wider bin would decrease the
contrast of the cluster counts, while a narrower bin would increase
the statistical uncertainty of the detection because of the lower
number of cluster counts.

In order to further justify our bin size, we build simulated clusters
at different redshifts ($z = $0.3, 0.5 and 0.8 respectively).  Our
simulated clusters have a projected density profile with a slope
$r^{-1}$ outside a core radius of 0.1 Mpc $\,h^{-1}$ and out
to a cut-off radius of 1 $\,h^{-1}$ Mpc. This choice is motivated by
the fact that PDCS clusters have, on average, a slope $r^{-1.4}$
(Lubin \& Postman 1996) and that steeper slopes produce clusters that
are more easily detected by the VGCF. We simulate the magnitudes of
cluster galaxies by sampling a Schechter luminosity function with a
slope $\alpha = -1.2$ and an absolute characteristic magnitude
$M_{V}^{*}=-20$. We apply both evolutionary and K-corrections to
cluster galaxies according to, respectively, Pence (1976) and
Poggianti (1996). We also take into account selection effects in the
visibility of galaxies (Phillips et al. 1990). We set the richness of
our simulated clusters according to the P96 definition of the richness
parameter $N_{R}$ (P96 clusters typically vary from $N_{R}=10$ to
$N_{R}=90$). As an example, in Fig. 5, we plot both the counts
within the area of a simulated cluster at z=0.3 (radius of 1 Mpc
$\,h^{-1}$ and $N_{R}=60$) and the average counts of the PDCS field
expected within the same area.  From the count distribution of our
simulated clusters, we find that a bin size of two magnitudes is
indeed adequate to contain the peak of counts corresponding to
clusters within the redshift range $z \sim 0.3\,-\,0.6$.

In the case of real clusters, the position of the peak of counts in
the magnitude distribution is not known \textit{a priori}.  Therefore,
we ``slide'' at small steps the magnitude bin over the whole magnitude
range of the catalog and run the VGCF at each step. We set the step to
0.1 magnitudes, the minimum possible step considering the typical
uncertainties of the PDCS magnitudes.

For simplicity, we start the VGCF with the bin $18.0 \leq V_{4} <
20.0$ and stop at the faint end of the survey, i.e. with the bin $21.8
\leq V_{4} < 23.8$. We neglect 76 galaxies brighter than $V_{4} =
18.0$ without consequences for our identification of clusters within
the redshift range $z\sim 0.3\,-\,0.6$.

\subsection{Clusters and their detections in bins}

Before producing our final cluster catalog, we need to define a
criterium to identify clusters by associating fluctuations detected in
different magnitude bins. The criterium will consists of a maximum
projected distance between the centers of fluctuations to be
associated and of a minimum number of coincident fluctuations required
for a positive detection, $N_{min}$.  In fact, because of the statistical noise
of the foreground/background galaxy distribution, the centers of the
fluctuations produced by a cluster in different magnitude bins will be
slightly different. As far as the number of coincident fluctuations
produced by a cluster is concerned, it will depend on the cluster
distance, richness and luminosity function.  Clearly the choice of the
criterium has to be made having in mind the goal of the detection
algorithm and the characteristics of the galaxy catalog.
\begin{table*}[!t]
\caption[ ]{The cluster catalog}
\begin{tabular}{|r|c|c|c|c|r|r|c|c|} 
\hline 
\hline 
N & A.R. (deg.) & Decl. (deg.)& R (arcsec.)  & Contrast & Ncl  & Nbg & Mag. bin & Remarks\\ 
\hline
  V1  &   201.623 &  29.977 &   227   &  7.91 &  25 &  10 &  4 (19.4) & P55\\
  V2  &   201.546 &  29.412 &   162   &  7.27 &  23 &  10 &  9 (19.9) & P49\\
  V3  &   201.437 &  29.492 &   133   &  4.47 &  10 &   5 & 11 (20.1) &  - \\
  V4  &   201.595 &  29.559 &   216   &  9.22 &  69 &  56 & 31 (22.1) & P50\\
  V5  &   201.695 &  30.043 &   112   &  5.50 &  11 &   4 & 13 (20.3) &  - \\
  V6  &   201.351 &  30.291 &   144   &  5.37 &  12 &   5 &  9 (19.9) &  - \\
  V7  &   201.095 &  29.783 &    58   &  6.36 &   9 &   2 & 30 (22.0) &  - \\
  V8  &   200.931 &  29.592 &   162   &  7.51 &  39 &  27 & 24 (21.4) & P51\\
  V9  &   201.381 &  29.640 &   137   &  9.65 &  32 &  11 & 26 (21.6) & P52\\
 V10  &   200.915 &  30.376 &    90   & 10.67 &  32 &   9 & 27 (21.7) & P62\\
 V11  &   201.498 &  30.365 &   155   &  5.81 &  26 &  20 & 21 (21.1) &  - \\
 V12  &   201.247 &  29.720 &   115   &  6.01 &  19 &  10 & 22 (21.2) & P53\\
 V13  &   201.549 &  29.928 &    61   &  6.00 &   6 &   1 & 21 (21.1) &  - \\
 V14  &   201.427 &  29.569 &    65   &  7.00 &   7 &   1 & 14 (20.4) &  - \\
 V15  &   201.079 &  29.918 &   151   &  4.67 &  14 &   9 & 21 (21.1) &  - \\
 V16  &   201.842 &  29.382 &    61   &  7.00 &   7 &   1 & 19 (20.9) & P46\\
 V17  &   201.910 &  30.094 &   101   &  6.64 &  23 &  12 & 28 (21.8) & P58\\
 V18  &   201.636 &  29.450 &   176   &  8.95 &  31 &  12 & 22 (21.2) &  - \\
 V19  &   201.603 &  29.845 &   115   &  6.00 &  12 &   4 & 23 (21.3) &  - \\
 V20  &   201.515 &  29.649 &   122   &  5.37 &  24 &  20 & 27 (21.7) &  - \\
 V21  &   201.278 &  30.409 &   104   &  6.15 &  23 &  14 & 30 (22.0) & P15\\
 V22  &   201.733 &  30.348 &   112   &  7.25 &  29 &  16 & 27 (21.7) &  - \\
 V23  &   200.935 &  29.678 &   108   &  6.50 &  13 &   4 & 22 (21.2) &  - \\
 V24  &   200.957 &  29.451 &   148   &  6.15 &  23 &  14 & 22 (21.2) &  - \\
 V25  &   201.836 &  29.856 &    68   &  6.05 &  16 &   7 & 34 (22.4) &  - \\
 V26  &   201.941 &  30.184 &   115   &  6.10 &  22 &  13 & 29 (21.9) &  - \\
 V27  &   201.942 &  29.550 &    72   &  6.35 &  11 &   3 & 26 (21.6) &  - \\
 V28  &   201.776 &  30.005 &    72   &  5.77 &  10 &   3 & 28 (21.8) & P56\\
 V29  &   201.935 &  29.500 &    79   &  5.50 &  11 &   4 & 29 (21.9) &  - \\
 V30  &   201.469 &  30.218 &   122   &  5.72 &  14 &   6 & 30 (22.0) &  - \\
 V31  &   200.959 &  29.359 &    97   &  5.55 &  20 &  13 & 32 (22.2) & P47\\
 V32  &   202.105 &  30.145 &   115   &  6.01 &  19 &  10 & 28 (21.8) &  - \\
 V33  &   201.866 &  30.140 &    68   &  5.00 &  10 &   4 & 28 (21.8) &  - \\
 V34  &   200.941 &  30.057 &    86   &  5.81 &  13 &   5 & 29 (21.9) & P57\\
 V35  &   200.906 &  30.080 &    68   &  6.00 &  12 &   4 & 38 (22.8) &  - \\
 V36  &   201.091 &  30.211 &    83   &  6.33 &  21 &  11 & 37 (22.7) & P63\\
 V37  &   200.983 &  30.108 &   111   &  5.63 &  27 &  23 & 38 (22.8) &  - \\
\hline
\end{tabular} 
\end{table*}
\begin{figure*}
\psfig{figure=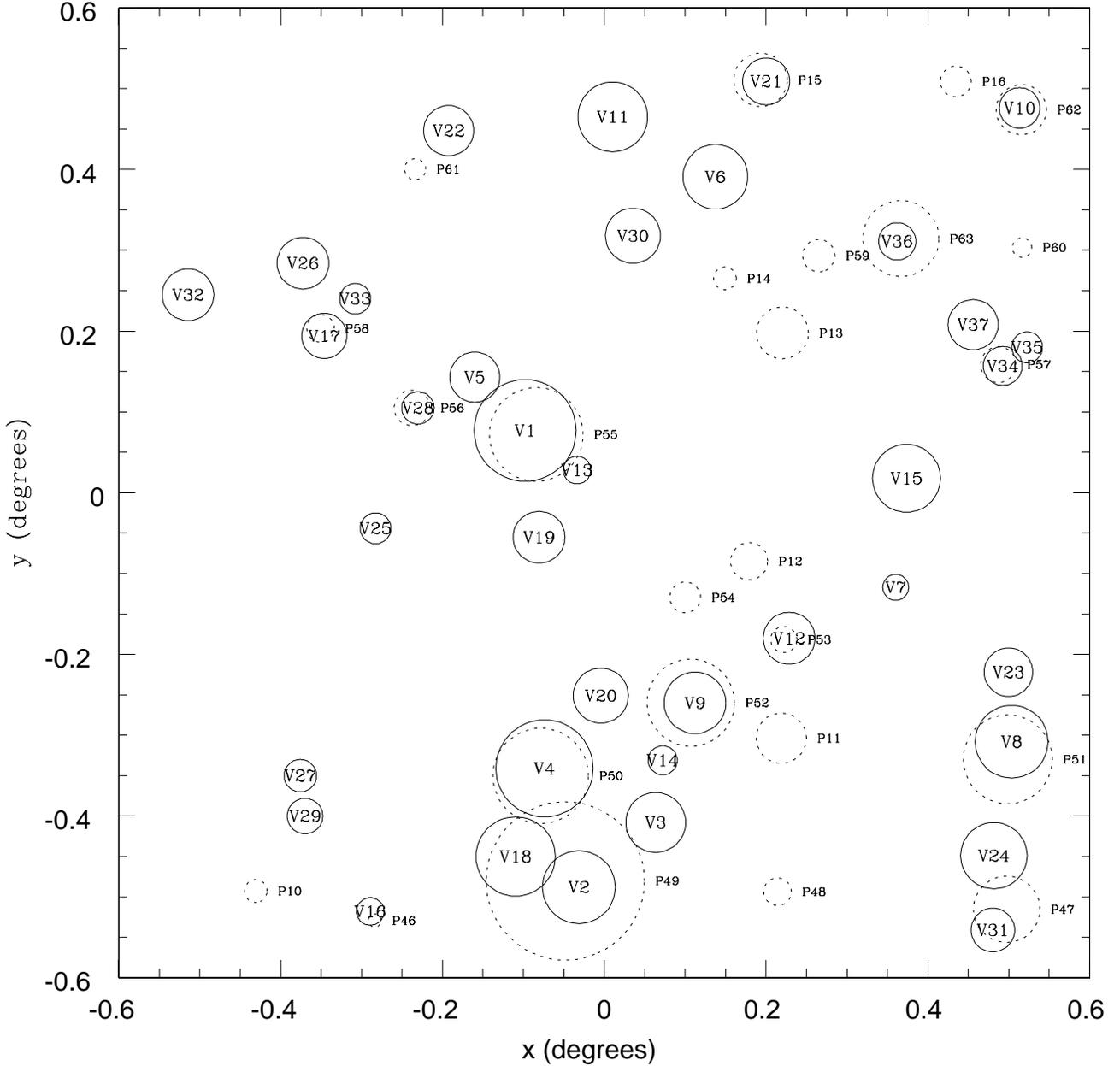,width=18cm,angle=0}
\caption{Clusters detected by the VGCF (solid circles) and by by the
matched filter algorithm (dotted circles).  VGCF identification
numbers are at the center of the solid circles, PDCS identification
numbers are on the side of dotted circles.  The field is centered at
$\alpha=$ 201.5 degrees and $\delta=$ 29.9 degrees (J2000).  North is
on the top and East on the left of the figure.}
%\label{fig8}
\end{figure*}
\begin{figure*}[t]
\psfig{figure=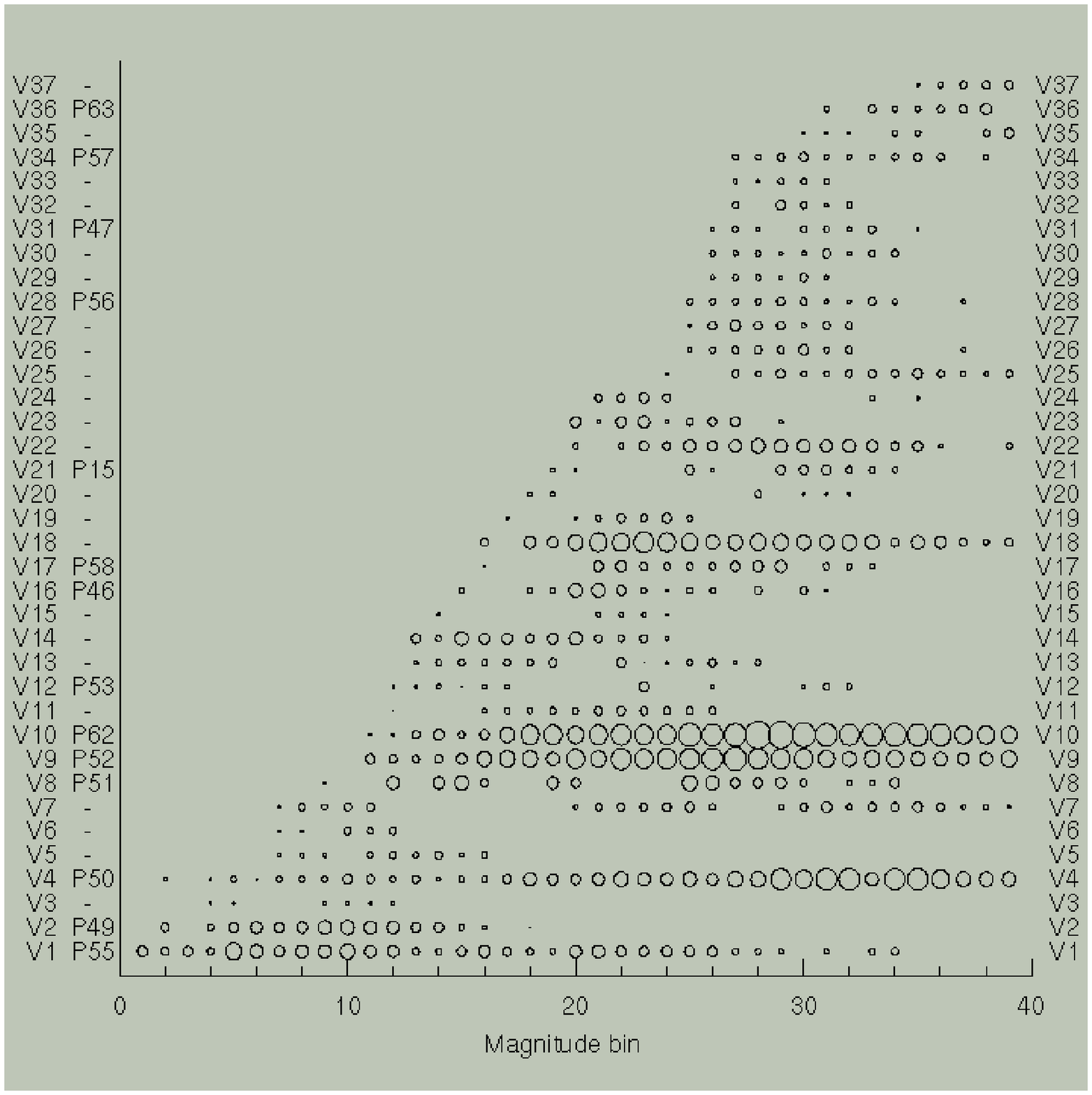,width=18cm,angle=0}
\caption{Plot of the density contrast of clusters detected by VGCF in
39 magnitude bins. Radii are scaled with the signal-to-noise ratio of
the detection. The labels on the left of the figure are the VGCF and
the PDCS identification numbers respectively.}
%\label{fig9}
\end{figure*}

In order to define the criterium of association for our test
application, we peform extended tests both on the PDCS field and on
simulations of Poissonian fields with embedded simulated clusters.
Our simulated fields have the same general properties of the PDCS field. 
In particular, we run the VGCF on 100 catalogs each containing 18 simulated 
clusters (a number similar to the number of clusters in the PDCS field) 
embedded within a Poissonian galaxy field. Clusters are simulated as in 
the previous subsection. All simulated catalogs contain 25432 galaxies, 
i.e. the same number of PDCS galaxies.

As a result of our tests, we consider coincident two fluctuations with
centers separated on the sky by a projected distance $d_{12} \leq
0.3\,min(R_1,R_2)$, where $R_1$ and $R_2$ are the radii of the two
fluctuations. A tighter criterium would break the sequence of
fluctuations corresponding to a real cluster, a looser criterium would
incorrectly associate fluctuations produced by adjacent
clusters/fluctuations to the same cluster.

We now set the minimum number of fluctuations, $N_{min}$, required for
the detection of a cluster. In Fig. 6 we plot the average number of
spurious fluctuations mis-identified as clusters as a function of
$N_{min}$. The number of spurious clusters drops dramatically as
$N_{min}$ increases from 1 to 5. For $N_{min} = 5$ there are on
average 1.5 spurious clusters per field. This number decreases slowly
as $N_{min}$ increases further. This result indicates that $N_{min} =
5$ is a a good choice to keep the number of Poisson fluctuations low
while still being sensitive to poor or distant clusters.

In Fig. 7 we show the variation with $N_{min}$ of the detection
efficiency of simulated clusters at redshifts z=0.3 (panel {\it a}),
0.5 (panel {\it b}), and 0.8 (panel {\it c}).  At each redshift the
different curves correspond to richnesses ranging from $N_R=10$ to
$N_{R}=60$. Clearly the fraction of detected clusters decreases as
$N_{min}$ increases.

We note that the curve in Fig. 6 corresponds to an evaluation of the
False Positive Rate (FPR), i.e. the probability of detecting as a
cluster a random fluctuation of the galaxy distribution. For
$N_{min}=5$, our FPR is very low. By comparison, P96 give 4.2 spurious
detections per square degree with peak signal greater than
$3\sigma$. The curves in Fig. 7 correspond to a measure of the False
Negative Rate (FNR).

\subsection{The VGCF run and comparison with P96}

We present here the results of the run of the VGCF on the $V_4$
catalog of the PDCS field at $\alpha = 13^{h}$ 26$^{m}$ and $\delta$ =
+$29^{o}$ 52' (J2000).  As discussed in the previous section, we run
the VGCF in bins two magnitude wide, ``sliding'' with 0.1 magnitude
steps within the magnitude range $18.00 \leq V_{4} < 23.8$.  In total
we run the VGCF in 39 magnitude bins and identify as clusters at least
five fluctuations that, according to our criterium, are angularly
coincident (see previous subsection).

On a Sun ULTRASparc 30 workstation, the time required to run the whole
procedure is about 40 minutes.

Our output cluster list consists of 37 objects. We characterize each
cluster with the properties of the fluctuation with the highest
signal-to-noise ratio as estimated by the ratio of the number of
cluster galaxies to the square root of the number of background
galaxies expected within the cluster area.
\begin{figure}[t]
\psfig{figure=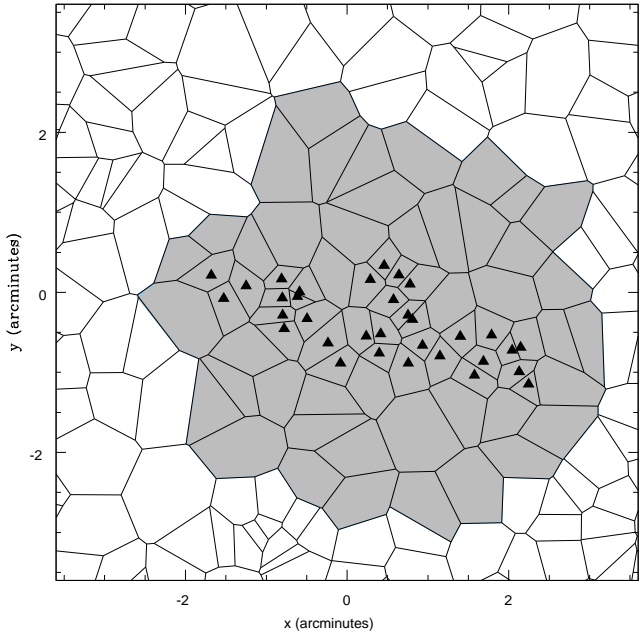,width=9cm,angle=0}
\caption{Plot of the galaxies of cluster V26 detected by the VGCF in
the bin of maximum contrast before regularization processes (black
triangles). The grey cells are those of galaxies added at the end of
the regularization processes. Empty cells are those of background
galaxies. North is on the top and East on the left of the figure.}
%\label{fig10}
\end{figure}

\begin{figure}[t]
\psfig{figure=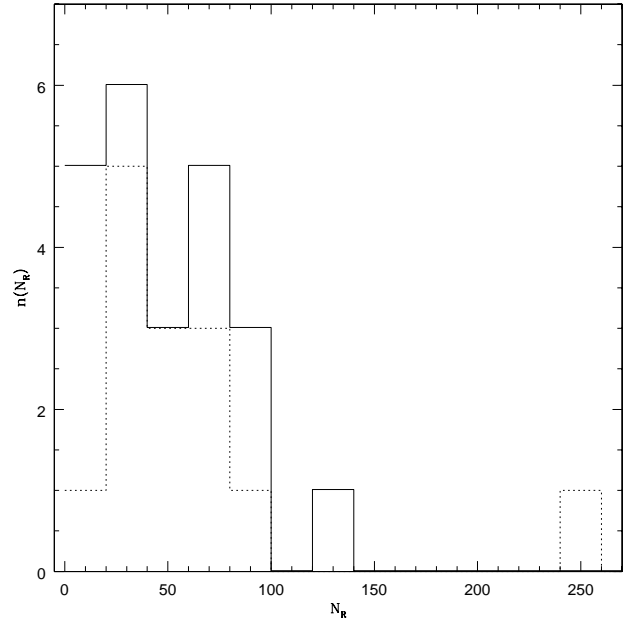,width=9cm,angle=0}
\caption{Richness distribution of P96 clusters as detected by the VGCF
(dotted line) and of new VGCF clusters (solid line).}
%\label{fig11}
\end{figure}
In Table 1, for each cluster we list: 1) identification number, 2)
J2000 right ascension and 3) J2000 declination, 4) radius, 5) the 
cluster signal-to-noise 
ratio, 6) the estimated number of cluster galaxies and 7) the number
of expected background galaxies, 8) the central magnitude of the bin
where we detect the cluster with the highest signal-to-noise ratio, 9)
the cross-identification with the PDCS catalog.

In Fig. 8 we plot circles on the sky corresponding to our clusters
(solid lines). We label our clusters with a ``V'' followed by their
order number in Table 1. In Fig. 9, we give a graphic summary of all
the fluctuations of each cluster. The abscissa is the order number of
the magnitude bin of the fluctuation and each cluster is represented
by a row of circles parallel to the magnitude bin axis. The radii of
the circles are scaled with the signal-to-noise ratio of the
detection.

We remind here that we fit a circle to a fluctuation after its 
detection, and that the only use of the circle is to provide a
convenient way to catalog the cluster with a center and a radius.

The richest and more reliable clusters exhibit, in Fig. 9, a
sequence of fluctuations. The signal-to-noise ratio of these
fluctuations regularly increases up to a maximum and then
decreases. Several fainter clusters show the same behavior, although
at a generally lower S/N level.  Clearly, the position of the
fluctuations along the magnitude axis is related to the cluster
distance.

Some clusters, for example V7 and V12, display substantial gaps along
the sequence. The suspicion is that these clusters may actually
consist of unrelated overdensities projected along the same
line-of-sight.

We now proceed to a detailed comparison with P96 clusters.  The
original PDCS catalog consists of 19 clusters detected by the matched
filter algorithm both in the $V_{4}$ band and in the $I_{4}$ band (see
P96, Table 4). Furthermore, the PDCS catalog lists 7 clusters detected
only in the $V_{4}$ band and added to the catalog after visual
inspection of the CCD images (see P96, Table 5).  In total there are
26 P96 $V_4$-band clusters, 18 of which are detected at the 3$\sigma$
level.  Of these last clusters, 13 have estimated redshift within the
range where our VGCF is optimal ($0.2 \leq z \leq 0.6$).  In Fig. 8
we plot circles corresponding to all the P96 clusters detected by the
matched filter technique (dotted line).

We identify 12 of the 13 clusters detected by P96 at the 3$\sigma$
level and with estimated redshifts from $z=0.2$ to $z=0.6$.  The only
cluster we do not identify, P54 ($z=0.5$), falls just under the
limit of our selection criterium with only four coincident
fluctuations.

Beside the above 12 clusters, we detect other 25 galaxy systems.  Of
these 25 systems, 2 are P96 clusters with confidence level less than
3$\sigma$ and redshift $z \leq 0.6$. The remaining 23 do not have a
counterpart in the PDCS catalog.

Most of the 23 new clusters have properties comparable to those of P96
clusters as detected by the VGCF. At least part of the new systems
have been detected by the VGCF thanks to its specific properties. For
example, because the VGCF does not smooth the data, it is able to
separate two large overdensities (V2 and V18) that are unlikely to be
one system (P49). In fact V2 (which we identify as the counterpart of
P49) and V18 are well separated both on the sky and in magnitude.

The property of the VGCF of detecting overdensities irrespective of
their shape is probably the key for the detection of V26, a very
elongated system. In Fig. 10 we show the galaxies of V26 as detected
by the VGCF before the regularization processes (black triangles). We
also show as grey tessels those of galaxies added at the end of the
regularization processes and as white tessels those of background
galaxies. The plot refers to the tessellation of the PDCS field in the
magnitude bin of maximum contrast for V26.

As far as the richness, $N_{R}$, of VGCF clusters is concerned, we
compute it almost in the same way as P96, the only difference being
that we use a fixed angular radius of 0.1 deg within which to count
cluster galaxies. For systems within the redshift range 0.3-0.6, our
angular radius roughly corresponds to a linear radius of 1 $\,h^{-1}$
Mpc. As P96 shows, $N_R$ is related to the Abell richness of the
system.

In Fig. 11 we plot the distribution of $N_R$ of our new systems
(solid histogram) and of P96 clusters as detected by the VGCF (dotted
histogram).  In general the richness of the new systems is similar to
that of P96 clusters (as detected by the VGCF), but the VGCF detects a
larger fraction of poor systems than the matched filter algorithm.  In
fact, we have 5 systems with $N_R < 20$ while only 1 P96 cluster falls
within this range. We probably find these poorer systems because we do
not smooth the data.

That the general properties of new VGCF systems are not very different
from P96 systems is also evident from Fig. 11 where we label with a
``P'' and the PDCS identification number the VGCF identifications of
P96 clusters.

A more detailed comparison between our and P96 clusters is not useful
since, in the end, only spectroscopic observations will allow a real
comparison between the performances of the two algorithms. What we
would like to stress here is that the VGCF is an algorithm that can
usefully complement the matched filter (and probably other parametric
cluster searches) for the definition of a complete and reliable 2D
cluster catalog.

In many problems we think that the VGCF could be preferred over
parametric searches and, in these cases, the example we discuss in
this section shows that the VGCF retrieves most of the reliable
identifications of the parametric searches.

% Section 4

\section{Summary}

We present an objective and automated procedure for detecting clusters
of galaxies in imaging galaxy surveys. Our Voronoi Galaxy Cluster
Finder (VGCF) uses galaxy positions and magnitudes to find clusters
and determine their main features: size, richness and contrast above
the background.  The VGCF uses the Voronoi tessellation to evaluate
the local density and to identify clusters as significative density
fluctuations above the background. The significance threshold needs to
be set by the user, but experimenting with different choices is very
easy since it only requires a new selection from the output of the
Voronoi tessellation and not a whole new run of the algorithm. The
VGCF is non-parametric and does not require a smoothing of the data.
As a consequence, clusters are identified irrispective of their shape
and their identification is only slightly affected by border effects
and by holes in the galaxy distribution on the sky.

The algorithm is fast, and automatically assigns members to
structures.  For example, a run on about 25000 galaxies only requires
7 minutes on a Sun ULTRASparc 30 workstation.

We perform test runs of the VGCF both on simulated galaxy fields and
on the 25432 galaxies identified in the $V_{4}$ band image of the PDCS
field centered at $\alpha = 13^{h}$ 26$^{m}$ and $\delta$ = +$29^{o}$
52' (J2000, see P96).  Given the depth of the PDCS survey, we run the
VGCF in 39 overlapping 2 magnitude wide bins, from $V_4 = 18.00$ to
$V_4 = 23.75$.  For the detection of a cluster we require at least
five angularly coincident fluctuations. Based on 100 simulations of
the PDCS field, we find that there are, on average, 1.5 spurious
detections in a simulated PDCS field with Poisson background. Our
choice of bin size and number of bins, together with the minimum
number of fluctuations required for a cluster detection, optimizes the
VGCF for the redshift range $0.3 \lesssim z \lesssim 0.6$.

We find 37 clusters, 12 of which are VGCF counterparts of the 13 PDCS
clusters detected at the 3$\sigma$ level and with estimated redshifts
from $z=0.2$ to $z=0.6$.  Of the remaining 25 systems, 2 are P96
clusters with confidence level $< 3\sigma$ and redshift $z \leq
0.6$.

23 VGCF clusters have no counterpart in the PDCS
catalog.  An inspection of their properties indicates that some of
these clusters may have been missed by the matched filter algorithm
for one or more of the following reasons: a) they are very poor, b)
they are extremely elongated, c) they lie too close to a rich and/or
low redshift cluster.

In conclusion, the VGCF can usefully complement the matched filter
algorithm (and probably other parametric cluster searches) for the
definition of a complete and reliable 2D cluster catalog.  In fact, we
think that the VGCF could be preferred over parametric searches in
many cases since the example we discuss shows that the VGCF anyway
retrieves the reliable identifications of parametric searches. 

% Section 5

\begin{acknowledgements}

We wish to thank Marc Postman for the galaxy catalog of the PDCS
field we used to prepare this paper.

\end{acknowledgements}

{}

\end{document}